\documentstyle[prl,aps,eqsecnum]{revtex}
\newcommand{\beq}{\begin{equation}}
\newcommand{\eeq}{\end{equation}}
\newcommand{\beqa}{\begin{eqnarray}}
\newcommand{\eeqa}{\end{eqnarray}}
\begin{document}
\input epsf.tex
\title{Gaussian Wigner distributions and hierarchies of
nonclassical states in quantum optics-The single mode case}
\author{Arvind\cite{email}} 
\address{Department of Physics\\
Indian Institute of Science,  Bangalore - 560 012, India}
\author{N. Mukunda\cite{jncasr}} 
\address{Center for Theoretical Studies and Department of Physics\\
Indian Institute of Science,  Bangalore - 560 012, India}
\author{R. Simon}
\address{Institute of Mathematical Sciences\\
C. I. T. Campus, Madras - 600 113, India}
\date{\today}
\draft
\maketitle
\maketitle
\begin{abstract}
A recently introduced hierarchy of states of a single mode quantised
radiation field is examined for the case of centered Guassian Wigner
distributions. It is found that the onset of squeezing among such states
signals the transition to the strongly nonclassical regime. Interesting
consequences for the photon number distribution, and explicit representations
for them, are presented.
\end{abstract}
\section{Introduction}
Squeezed states of light, and other states exhibiting either
antibunching or subpoissonian photon statistics or both, are
well known examples of so called ``nonclassical'' states of
radiation~\cite{walls-nature}~\cite{squeezing-1}~\cite{antibunching}~
\cite{sub-poissonian}. 
In fact these are the simplest and most familiar ones
out of an infinite hierarchy of independent signatures of
nonclassical states in quantum optics; many other signatures have
been presented in the literature~\cite{others}.

The precise definition of a nonclassical state of radiation is
based upon the diagonal coherent state expansion of the density
matrix $\hat{\rho}\/$ of the state in the quantum theory.
Limiting ourselves to the single mode radiation field this expansion
is~\cite{diag-coh}
\beq
\hat{\rho}=\int\frac{d\textstyle^2z}{\textstyle \pi} \phi(z) \vert z
\rangle \langle z \vert \; ,
\label{diag-coh-def}
\eeq
where the coherent states $\vert z \rangle\/$ are the familiar
normalised eigenstates of the photon annihilation operator
$\hat{a}\/$ with complex eigenvalue $z\/$ and $\phi(z)\/$ is
a real normalised weight function which is in general a
distribution. The state $\hat{\rho}\/$ is said to be ``classical
'' if $\phi(z)\/$ is pointwise nonnegative, and nowhere more
singular than a delta function, so that it can be interpreted as
a classical probability density over the complex plane. Otherwise
$\hat{\rho}\/$ is a ``nonclassical '' state.
This classification is clearly invariant under 
rotations and translations in phase space.

It has been shown elsewhere that there is a dual operator based
approach to this distinction between classical and nonclassical
states, which is physically quite instructive~\cite{characterization}. The
representation~(\ref{diag-coh-def}), as is well known, is closely
related to the normal ordering rule of correspondence between
classical dynamical variables and quantum operators. Given any
real classical function $f(z^{\star},z)\/$ of a complex variable
$z\/$ and its conjugate, one defines a hermitian operator
$\hat{F}\/$ in quantum theory by the replacement $z \rightarrow
\hat{a}, z^{\star} \rightarrow \hat{a}^{\dagger}\/$ and then
bringing all factors $\hat{a}^{\dagger}\/$ ``by hand'' to the
left of all factors $\hat{a}\/$:
\beqa
f(z^{\star},z) \rightarrow \hat{F} &=& f(\hat{a}^{\dagger},
\hat{a}) \vert_{\mbox{ $\hat{a}^{\dagger}\/$ to left, 
$\hat{a}\/$ to right}}\/, \nonumber \\
\langle z \vert \hat{F} \vert z \rangle &=& f(z^{\star},z) .
\label{normal-order}
\eeqa 
Then the quantum mechanical expectation value of $\hat{F}\/$ in
the state $\hat{\rho}\/$ is 
\beq
\langle \hat{F} \rangle = {\rm Tr} (\hat{\rho}
\hat{F})=\int\frac{d\textstyle^2z}{\textstyle \pi} \phi(z) 
f(z^{\star},z).
\eeq
The key observation now is that while the correspondence $f\leftrightarrow
\hat{F}\/$ is linear and takes real functions to hermitian
operators and vice versa, a real nonnegative $f(z^{\star},z)\/$
may well lead to a hermitian indefinite $\hat{F}\/$. A state
$\hat{\rho}\/ $ is then said to be classical if this permitted
`` quantum negativity '' in operators never shows up in
expectation values, nonclassical otherwise:
\beqa
\hat{\rho}\quad {\rm Classical} &\Leftrightarrow& {\rm Tr}(\hat{\rho}
\hat{F}) \geq 0 \;\;\mbox{for every} \; f(z^{\star},z)  \geq 0,
\nonumber \\
\hat{\rho}\quad {\rm Nonclassical} &\Leftrightarrow& {\rm Tr}(\hat{\rho}
\hat{F}) < 0 \;\;\mbox{for some} \; f(z^{\star},z)  \geq 0.
\label{nonc-op}
\eeqa

With this alternative characterization (completely equivalent to the
usual one), one has the possibility of defining several degrees or
levels of nonclassicality, if one restricts in various ways the
collection of operators $\hat{F}\/$ for which one tests the
conditions given in~(\ref{nonc-op})~\cite{characterization}.
Specifically, for a single mode system, it has been shown by
considering the subset of phase invariant (number conserving)
operators $\hat{F}\/$ which arise from $f(z^{\star},z)\/$ obeying
\beq
f(z^{\star} e^{-i \alpha},z e^{i \alpha})=f(z^{\star},z)
\label{phase-inv-funct}
\eeq
that an exhaustive and mutually exclusive three-fold classification
of states is possible. If $f(z^{\star},z)\/$
obeys~(\ref{phase-inv-funct}), then for the expectation value
of the corresponding $\hat{F}\/$ it suffices to use an angle
average of $\phi(z)\/$:
\beqa
[\hat{F}, \hat{a}^{\dagger} \hat{a}]&=&0 \Rightarrow {\rm
Tr}(\hat{\rho} \hat{F}) = \int \limits^{\infty}_{0} dI {\cal
P}(I)\;f(I^{1/2},I^{1/2}),\nonumber \\
{\cal P}(I) &=& \int \limits_0^{2 \pi} \frac{\displaystyle d
\theta}{\displaystyle 2 \pi} \phi(I^{1/2} e^{i \theta}) .
\eeqa  
One can then obtain the following finer classification of all
states: 
\beqa
\hat{\rho}\quad {\rm Classical} &\Leftrightarrow& 
\phi(z) \geq 0, \;\;\mbox{so} \; {\cal P}(I)  \geq 0,\nonumber \\
\hat{\rho}\quad \mbox{Weakly nonclassical} &\Leftrightarrow& 
{\cal P}(I) \geq 0,\;\mbox{but}\; \phi(z) \not \geq 0 \nonumber \\
\hat{\rho}\quad \mbox{Strongly nonclassical} &\Leftrightarrow& 
  {\cal P}(I)  \not \geq 0,\;\mbox{so}\; \phi(z) \not \geq 0 
\label{three-fold}
\eeqa
Thus the previous ``nonclassical'' has been subdivided now into
``weakly nonclassical`` and ``strongly nonclassical'' states. Upto
and including the weakly nonclassical level, ${\cal P}(I)\/$ can
be treated as a classical probability density for intensity,
whether or not $\phi(z)\/$ can be regarded as a probability
distribution over the complex plane; in the third strongly
nonclassical regime, even ${\cal P}(I)\/$ ceases to be a
probability density.

The aim of this paper is to illustrate these ideas in the
concrete case of states described by Gaussian-Wigner
distributions on phase space. It is well known that in a wide
variety of physical processes the states of radiation that are
produced are indeed of this type~\cite{process-gaussian}. 
Their description also lends
itself to direct analytical treatment. What we shall
demonstrate is that within this set of states, the onset of
squeezing signals an abrupt change from classical to the
strongly nonclassical regime; thus the weakly nonclassical
states do not show up at all in this family!

The material of this paper is arranged as follows. In Section II
we trace the connection between the descriptions of an operator
via its
Weyl weight and its Wigner representative, and the diagonal weight
$\phi(z)\/$. This gives us a clear picture of the extent to
which $\phi(z)\/$ can be a singular distribution, and in turn
how singular the quantity ${\cal P}(I)\/$ can in principle be.
Section~III examines the class of centered Gaussian Wigner
distributions. These are fully parametrised by the variance or
noise matrix which has to be positive semidefinite and also
must obey the uncertainty principle. Among these states the only
two qualitatively different ones are the nonsqueezed and
squeezed ones. In the former case, both $\phi(z)\/$ and ${\cal
P}(I)\/$ can be computed explicitly, and as expected they are
finite nonnegative normalized functions. This is consistent with
their being classified as classical states. In contrast, the
squeezed states are shown to be strongly nonclassical, and one
never sees the weakly nonclassical possibility at all. Section~IV
gives an example of weakly nonclassical states which are
naturally outside
the Gaussian Wigner family, and offers some
concluding remarks.
\section{Nature of the Distributions $\phi(\protect\lowercase{z})\/$ and 
${\cal P}(I)\/$ } 
It is useful to begin by recalling the general properties of the
diagonal weight $\phi(z)\/$ and its angular average ${\cal
P}(I)\/$, and by giving an indication of the kinds of singular
distributions we must be prepared to
encounter~\cite{distributions}. This is best
done by viewing the set of all possible density matrices
$\hat{\rho}\/$ as a subset of the family of Hilbert-Schmidt
(H-S) operators. An operator $A\/$ on Hilbert space is of H-S
type if 
\beq
\mbox{Tr}(A^{\dagger}A) < \infty,
\label{hilbert-shmidt}
\eeq   
and among H-S operators we have  a natural inner product :
\beq
(A,B) = \mbox{Tr} (A^{\dagger} B).
\eeq
We deal throughout with  systems of one degree of freedom, and with 
the annihilation and creation operators $\hat{a}, \hat{a}^{\dagger}\/$ 
related to hermitian  $\hat{q}\/$ and $\hat{p}\/$ in the standard
way:
\beq
\hat{a} = 
\frac{\displaystyle 1}{\displaystyle \sqrt{2}}(\hat{q}+i\hat{p}),\quad
\hat{a}^{\dagger} = \frac{\displaystyle 1}
{\displaystyle \sqrt{2}}(\hat{q}-i\hat{p})
\eeq
The unitary phase space displacement operators are defined by and have the
following properties:
\beqa
D(\sigma, \tau)
 &=& \mbox{exp}(i\sigma \hat{q} - i \tau \hat{p}), \quad -\infty
< \sigma, \tau < \infty; \nonumber \\
D(\sigma, \tau)^{\dagger}&=&
D(\sigma, \tau)^{-1} =
D(-\sigma, -\tau) ; \nonumber \\
\mbox{Tr} \left(D(\sigma^{\prime}, \tau^{\prime})^{\dagger}
  D(\sigma, \tau)\right) &=& 2 \pi \delta(\sigma^{\prime}-\sigma)
\delta(\tau^{\prime}-\tau).
\eeqa
Any H-S operator $A\/$ can be expanded in the form of an 
operator Fourier integral representation using its ``Weyl weight'' 
$\tilde{A}(\sigma, \tau)\/$ as expansion coefficient~\cite{weyl-book}:
\beqa
A&=&\int \int \frac{\displaystyle d \sigma d \tau}{\displaystyle 
\sqrt{2 \pi}} \tilde{A}(\sigma, \tau) D(\sigma, \tau), 
\nonumber \\
\tilde{A}(\sigma, \tau)&=&\frac{\displaystyle 1}
{\displaystyle \sqrt{2 \pi}} 
\left( D(\sigma, \tau), A \right),
\nonumber \\
\mbox{Tr}(A^{\dagger}A)&=&(A,A)=\int \int d \sigma \; d\tau \vert
\tilde A(\sigma, \tau) \vert^{2}.
\label{weyl-weight}
\eeqa
Thus the H-S property~(\ref{hilbert-shmidt}) of $A\/$ is translated exactly
into the $L^{2}$ property of $\tilde{A}(\sigma, \tau)\/$ over
${\cal R}^{2}$.
\par
From $\tilde{A}(\sigma, \tau)\/$ we pass to the Wigner
representative or Wigner distribution $W(q,p)$ of the
operator $A$ by a double Fourier transform at the
c-number level~\cite{wigner-1932}:
\beq
W(q,p)=\int \int \frac{\displaystyle d \sigma d \tau}{\displaystyle 
(2 \pi)^{\frac{3}{2}}}\tilde{A}(\sigma, \tau) \mbox{exp}\left(
i \sigma q-i \tau p \right)
\label{wigner-weight}
\eeq
Here $q\/$ and $p\/$ are canonical coordinates over a classical
phase space, and in case $A\/$ is hermitian its Wigner
representative $W(q,p)\/$ is real. Now the H-S property for
$A\/$ amounts to $W(q,p)\/$ being an $L^{2}\/$ function over
${\cal R}^{2}\/$:
\beq
  Tr(A^{\dagger}\/A) = (A,A) = 2\pi\int\int dq\/ dp \vert{W(q,p)} \vert^{2}
\eeq
For density matrices we are also interested in the ordinary trace:
\beq
Tr(A) = \sqrt{2\pi}\tilde{A}(0,0) = \int \int dq\/dp W(q,p)
\eeq

It is in the passage from $\tilde{A}(\sigma, \tau)\/$ 
or $W(q,p)\/$ to $\phi(z)\/$ that the distribution character
of the latter shows up. From the diagonal representation
\beqa
A=\int \frac{\displaystyle dx\/dy}{\displaystyle 2\pi}\;
\phi(z)\; \vert z \rangle \langle z \vert,
\eeqa
where $z = \frac{1}{\sqrt{2}}(x+iy)\/$, when we connect up
with the previous
relations~(\ref{weyl-weight},~\ref{wigner-weight}) 
we get the result:
\beqa
\phi(z)&=&\int \int \frac{\displaystyle d \sigma d \tau}{\displaystyle 
\sqrt{2 \pi}}\/ e^{\frac{1}{4}(\sigma^2+\tau^2)} 
\tilde{A}(\sigma, \tau) e^{i(\sigma x -\tau y)} 
\nonumber\\
&=&\int \int \frac{\displaystyle d \sigma d \tau}{\displaystyle 
2 \pi}\/ e^{\frac{1}{4}(\sigma^2+\tau^2)+i(\sigma x - \tau
y)}\int \int dq \/dp W(q,p) e^{i\/(\tau p - \sigma q)}.
\label{diag-coh-wigner}
\eeqa
Thus the most singular kind of $\phi(z)$ is one whose
Fourier transform is the increasing Gaussian factor
$\exp{\frac{1}{4}(\sigma^{2}+\tau^{2})}$ times a square
integrable function $\tilde{A}(\sigma, \tau)\/$ - this is
the worst behaviour that can in principle occur. Conversely
for a classical state $\tilde{A}(\sigma, \tau)\/$ must more
than overwhelm this exponential factor and moreover yield a
nonnegative $\phi(z)$.
\par
Let us next see what this situation for $\phi(z)$ entails
for its angular average ${\cal P}(I)$. We work directly with the
Wigner distribution $W(q,p)$ and find after performing
the angular integration:
\beqa
{\cal P}(I) &=& \int \limits^{2 \pi}_{0} \frac{\displaystyle d
\theta} {\displaystyle 2 \pi} \phi(I^{\frac{1}{2}}\/e^{i
\theta}) \nonumber\\
&=&\int \int \frac{\displaystyle d \sigma d \tau}{\displaystyle 
2 \pi}\/ e^{\frac{1}{4}(\sigma^2+\tau^2)}
J_0(\sqrt{2 I (\sigma^2+\tau^2)})
\int \int dq \/dp W(q,p) e^{i\/(\tau p - \sigma q)}.
\eeqa
If we substitute $\sigma = \sqrt{2K}\cos{\psi}\/$, 
$\tau = \sqrt{2K}\sin{\psi}\/$, we can carry out one more
angular integration and bring ${\cal P}(I)\/$ to the following form:
\beqa
{\cal P}(I) &=&  \int \limits^{\infty}_{0}
dK e^{{\frac{K}{2}}}.\/ J_0(2 \sqrt{IK})
\int \int dq\/dp W(q,p)\;J_0(\sqrt{2\/ K\/ (q^2 + p^2})
\nonumber \\
&=&\int \limits^{\infty}_{0}
dK e^{{\frac{K}{2}}}.\/ J_0(2 \sqrt{IK})
\int \limits^{\infty}_{0} dL J_0(2 \sqrt{KL})
\int \limits^{2 \pi}_{0} d\chi\/ W(\sqrt{2 L} \cos{\chi},
\sqrt{2 L} \sin{\chi}) 
\label{PI-wigner}
\eeqa
Now just as the relation~(\ref{diag-coh-wigner}) between $\phi(z)\/$
and $\tilde{A}(\sigma, \tau)\/$ involved the classical
two dimensional Fourier transformation, here one is concerned
with the single variable Fourier-Bessel transformation over
the half-line $(0,\infty)\/$ which states~\cite{forier-bessel}:
\beqa
\int \limits^{\infty}_{0} dI \vert f(I)\vert^2 < \infty 
&\Rightarrow& \nonumber \\
f(I) &=& \int \limits^{\infty}_{0} d K g(K) J_0(2 \sqrt{IK}),
\nonumber \\
g(K) &=& \int \limits^{\infty}_{0} d I f(I) J_0(2 \sqrt{IK}),
\nonumber \\
\int \limits^{\infty}_{0} dI \vert f(I)\vert^2
&=&\int \limits^{\infty}_{0} dK \vert g(K)\vert^2,
\nonumber \\
\int \limits^{\infty}_{0} d K J_0(2 \sqrt{LK}) J_0(2 \sqrt{IK})
&=&\delta(I-L).
\label{fourier-bessel}
\eeqa
This means that the most singular possible behaviour for ${\cal
P}(I)\/$ 
which can in principle occur is that its Fourier-Bessel
transform can be the factor $e^{\frac{K}{2}}$ times a square
integrable function of $K\/$  over the domain $(0,\infty)\/$ , namely
the Fourier-Bessel transform of the angular average of
$W(q,p)$. The factor $e^{\frac{K}{2}}\/ $ 
is just the earlier factor $e^{\frac{1}{4}(\sigma^2 + \tau^2)}\/$ present
in eq~(\ref{diag-coh-wigner}); and the situation for ${\cal P}(I)\/$
 is marginally better than
for $\phi(z)\/$ since now only the angular average of $\phi(z)\/$
is involved.
\par
The use of phase space language in describing operators
in quantum mechanics leads naturally to an examination of
the behaviours of $\phi(z)$ and ${\cal P}(I)\/$ under phase space
rotations and translations. As is easy to see, their
behaviour under rotations is simple:
\beq
W^{\prime}(q,p) = W(q \cos{\alpha} - p \sin{\alpha}, p
\cos{\alpha} + q \sin{\alpha}) \Leftrightarrow
\phi^{\prime}(z)=\phi(z\/e^{i\alpha}) 
\Rightarrow {\cal P}^{\prime}(I) =  {\cal P}(I).
\eeq
This invariance of ${\cal P}(I)\/$ is as expected. Under translations
we have
\beq
W^{\prime}(q,p) = W(q-q_0 , p-p_0) \Leftrightarrow
\phi^{\prime}(z) = \phi(z-z_0),\quad z_0=\frac{1}{\sqrt{2}}(q_0+ip_0)
\eeq
However now ${\cal P}^{\prime}(I)$ is not expressible in terms
of ${\cal P}(I)\/$
alone as phase sensitivity is introduced by a translation.
Therefore while our threefold classification scheme~(\ref{three-fold})
is obviously invariant under phase space rotations, the
behaviour with respect to translations is much more subtle.

It is evident that the classical states with both $\phi(z)\/$
and ${\cal P}(I)\/$ nonnegative remain classical under
translations. However a weakly nonclassical state 
becomes strongly nonclassical for a suitably chosen translation,
as the following physical argument shows. 
At the origin ${\cal P}(0)\/$ reduces to $\phi(0)\/$
as no angular average remains.
If a weakly nonclassical state is given, its $\phi(z)\/$ must become
effectively negative somewhere in the complex plane. By translating
the origin to such a point and then computing ${\cal
P}^{\prime}(0)\/$ we see that the resulting state is strongly
nonclassical. Following a similar argument we also see that we can
recover $\phi(z)\/$ in its entirety by subjecting the initial state
to all possible phase space displacements $z_0\/$, 
$\phi^{\prime}(z)=\phi(z-z_0)\/$, and then computing the resulting
${\cal P}^{\prime}(I)\/$ and collecting the results.

We conclude this Section by relating the distribution ${\cal
P}(I)\/$ to the photon number probabilities. Indeed these
involve a complete independent set of phase insensitive
quantities and their expectation values:
\beqa
f(z^{\star},z) &=& {\displaystyle e}^{-z^{\star}\/z}\;
\frac{\displaystyle (z^{\star}\/z)^{n}}{\displaystyle n!}
\leftrightarrow \hat{F} = \vert n \rangle \langle n \vert ,
\nonumber\\ 
p(n) &=&{\rm Tr}(\hat{\rho} \hat{F}) =\langle n \vert \hat{\rho}
 \vert n\rangle \nonumber \\
&=&\int\limits^{\infty}_{0} dI\/ {\cal P}(I) {\displaystyle e}^{-I}
\;\frac{\displaystyle I^n}{\displaystyle n!}
\eeqa
These $p(n)$'s always give well defined normalised probabilities for
finding various numbers of photons, whether or not ${\cal P}(I)\/$ is
itself a probability density. Formally one can invert the above to get
${\cal P}(I)\/$ in terms of $p(n)\/$, as indeed one would expect. If
we define the generating function $q(K)\/$ by 
 \beq
q(K)= \sum^{\infty}_{n=0}\frac{\displaystyle (-1)^n}{n!}
K^n p(n)
\eeq
we see that $q(K)\/$ converges for all real $K\/$ and is related to
${\cal P}(I)\/$ by
\beqa
q(K)&=& \sum^{\infty}_{n=0}\frac{\displaystyle (-1)^n}{n!}
K^n \int \limits^{\infty}_{0}dI\/ {\cal P}(I) {\displaystyle e}^{-I}
\;\frac{\displaystyle I^n}{\displaystyle n!}
\nonumber \\
&=&\int \limits^{\infty}_{0}dI\/ {\cal P}(I) {\displaystyle e}^{-I}
J_0(2 \sqrt{IK}).
\eeqa
Using the formula~(\ref{fourier-bessel}) 
of the Fourier Bessel transformation again
we get the inversion
\beq
{\cal P}(I)= {\displaystyle e}^{I}\/
\int \limits^{\infty}_{0}dK\/ q(K) 
J_0(2 \sqrt{IK}).
\eeq
In the classical and weakly nonclassical cases, then, the generating 
function $q(K)\/$ is itself well behaved and leads to nonnegative
${\cal P}(I)\/$, but in the strongly nonclassical case, it causes
${\cal P}(I)\/$ to be a distribution, or at any rate not a
probability. 
\section{The case of Gaussian Wigner distributions}
We consider the family of centered Gaussian Wigner distributions, namely
those which have vanishing means for $q\/$ and
$p$~\cite{gaussian-wigner-1}. The most general such
distribution is determined by a real symmetric $2\times 2\/$ matrix $G\/$
\beqa
W_G(q,p) &=& \frac{\displaystyle \sqrt{\mbox{det} G}}{\displaystyle \pi}
\mbox{exp}\left(\begin{array}{c} \\ \end{array}-(q\quad p)\;G\;
 \left(\begin{array}{c} q \\ p \end{array}
\right) \right), \nonumber \\ 
G &=& \left( \begin{array}{cc} A & B \\ B & C \end{array} \right) .  
\eeqa
The condition that $W_G(q,p)\/$ represent a physically realisable
quantum mechanical state imposes the following restrictions on $G\/$ 
corresponding respectively to normalisability and the uncertainty 
principle~\cite{gaussian-wigner-2}:
\begin{mathletters}
\beqa
G > 0,\quad \mbox{ie}\quad A+C>0,\quad \Delta=\mbox{det}G=AC-B^2 >0;
\\
G^{-1}+i\left(\begin{array}{cc}0 & 1 \\ -1 & 0 \end{array} \right)
\quad \geq 0 \;, \; \mbox{ie} \quad A+C \geq 0\;, \quad \Delta \geq 
\Delta^2.
\eeqa
\end{mathletters}
Combining these we have the complete set of restrictions on $G\/$ given by
\beq
A+C > 0,\quad 0 < \Delta \leq 1.
\eeq
The noise or variance matrix $V\/$ is defined and given by
\beqa 
V&=&\left( \begin{array}{cc} (\Delta q)^2 & \Delta(q,p) \\
\Delta (q,p) & (\Delta p)^2 \end{array}\right)=
\frac{1}{2} G^{-1} =\frac{\displaystyle 1}{\displaystyle 2 \Delta}
\left( \begin{array}{cc}C&-B\\-B&A \end{array}\right) \nonumber \\
(\Delta q)^2 &=& \int \int  dq\; dp\;q^2\; W_G(q,p), \nonumber \\ 
\Delta(q,p) &=& \int \int dq\;  dp \; q p\;  W_G(q,p), \nonumber \\ 
(\Delta p)^2 &=& \int \int dq\; dp \;p^2\; W_G(q,p). 
\eeqa

Here the vanishing of the means of $q\/$ and $p\/$ has been used.
In terms of $V\/$, the uncertainty principle appears 
in the following form~\cite{squeezing}:
\beq 
\mbox{det} V = \frac{\displaystyle 1}{\displaystyle 4  \Delta} \geq 
\frac{\displaystyle 1}{\displaystyle 4}. 
\eeq

We can use the covariance of $\phi(z)\/$ and the invariance of ${\cal P}(I)
\/$ under phase space rotations to simplify the situation and to assume
without loss of generality that $G\/$ and $V\/$ are diagonal. Moreover 
these rotations do not disturb the three-fold classification of 
states~(\ref{three-fold}). Therefore we parametrise $G\/$ and $V\/$ using 
two real positive parameters $\alpha \/$ and $\beta\/$ as follows:
\beqa
V&=&\frac{\displaystyle 1}{\displaystyle 2} 
\left( \begin{array}{cc} \alpha^2 & 0\\
0 & \beta^2 \end{array}\right) \nonumber 	\\
G&=& 
\left( \begin{array}{cc} 1/ \alpha^2 & 0\\
0 & 1/ \beta^2 \end{array}\right),\; \alpha , \beta >0, \alpha
\beta \geq 1; \nonumber \\
W_{(\alpha, \beta)}(q,p) &=&
 \frac{\displaystyle 1}{\displaystyle \pi \alpha \beta}
 \mbox{exp}\left(
 -\frac{\displaystyle q^2}{\displaystyle \alpha^2 }
 -\frac{\displaystyle p^2}{\displaystyle \beta^2}
\right). 
\eeqa
To deal with $\phi(z)$ and ${\cal P}(I)$ we need respectively
the Fourier transform and the angular average of
$W_{(\alpha,\beta)}(q,p)$; these are:
\begin{mathletters}
\beqa
 \int\int\/dq dp W_{(\alpha,\beta)}(q,p) exp(i\tau p-i\sigma q)&=&
\mbox{exp}\left(-\frac{\alpha^2 \sigma^2}{4} -
\frac{\beta^2 \tau^2}{4} \right);
\label{wigner-gaussian-FT}
\\
\int\limits^{2 \pi}_{0} d \chi W_{(\alpha, \beta)}
(\sqrt{2L}\cos{\chi}, \sqrt{2 L} \sin{\chi})&=&
\frac{\displaystyle 2}{\displaystyle \alpha \beta}
\mbox{exp}\left(-L(\frac{\displaystyle 1}{\displaystyle
\alpha^2}+\frac{\displaystyle 1}{\displaystyle
\beta^2} ) \right) I_{0}\left( L(\frac{\displaystyle 1}
{\displaystyle \alpha^2} - \frac{\displaystyle 1}{\displaystyle
\beta^2})\right)
\label{wigner-angle-av}
\eeqa
\end{mathletters}
Here $I_{0}(w)=J_{0}(iw)\/$ is the Bessel function of order zero
and imaginary argument.
\par
Returning to the Wigner function $W_{(\alpha,\beta)}(q,p)$,
the nonsqueezed case corresponds to both $\alpha,\beta \geq 1$;
while if one of them becomes less than unity we have a squeezed
state. For definiteness in the latter case we take $p\/$ to be
the squeezed variable, so we take $\beta < 1\/$ and
$\alpha > 1\/$ maintaining $\alpha\beta \geq 1\/$. Formally
we have throughout, on combining eqs~(\ref{diag-coh-wigner}
,~\ref{wigner-gaussian-FT}):
\beq
\phi_{(\alpha, \beta)}(z)=\int \int 
\frac{\displaystyle d\sigma\/ d \tau}{\displaystyle 2 \pi}
{\displaystyle e}^{i(\sigma x - \tau y)}\/
\mbox{exp} \left[-\frac{1}{4}(\alpha^2-1)\sigma^2 -
\frac{1}{4}(\beta^2-1)\tau^2\right]
\label{diag-coh-gauss}
\eeq
In the nonsqueezed regime these integrals can be computed and
we get expected results:
\beqa
\phi_{(\alpha, \beta)}(z)=\left\{ \begin{array}{l}
2\/(\alpha^2-1)^{-1/2}.(\beta^2-1)^{-1/2}
\mbox{exp}\left[ -\frac{\displaystyle x^2}{\displaystyle
\alpha^2-1} -\frac{\displaystyle y^2}{\displaystyle
\beta^2-1} \right],\quad \alpha, \beta >1;\\ 
\sqrt{2 \pi}\delta(x)\/\sqrt{2}(\beta^2-1)^{-1/2}
\/\mbox{exp}\left( -\frac{\displaystyle y^2}{\displaystyle 
\beta^2-1}\right),
\quad \alpha =1, \beta >1;\\
\sqrt{2 \pi}\delta(y)\/\sqrt{2}(\alpha^2-1)^{-1/2}
\/\mbox{exp}\left(-\frac{\displaystyle x^2}{\displaystyle 
\alpha^2-1}\right),
\quad \alpha >1, \beta =1;\\
2\/ \pi \delta(x)\/ \delta(y)\/,\quad \quad \alpha=\beta=1.
\end{array} \right.
\eeqa
\par
In all these cases the state is classical. However, once
$\beta \/$ dips below unity, we see from eq~(\ref{diag-coh-gauss}) that the 
Fourier transform of $\phi(z)\/$ is an increasing Gaussian
in the variable $\tau\/$. This means that $\phi(z)\/$ has switched
abruptly to being a distribution, essentially of the most 
singular kind that can arise. (Of course, if
$\beta\/$ continually decreases and squeezing increases, $\phi(z)\/$
does become more and more singular). This is consistent
with squeezed states being nonclassical. The interesting point
is that there is no intermediate regime (among "Gaussian-Wigner"
states) in which the singularity of $\phi(z)\/$ is somewhat milder,
say involving finite order derivatives of delta functions.
\par
To follow the behaviour of ${\cal P}_{(\alpha,\beta)}(I)\/$ as we
pass from the nonsqueezed state to the squeezed regime,
and when $\beta < 1$ to discriminate between the weakly
nonclassical and the strongly nonclassical possibilities,
we begin by combining
eqs~(\ref{PI-wigner}~\ref{wigner-angle-av}) to get a formal
integral expression for ${\cal P}_{(\alpha,\beta)}(I)\/$:
\beq
{\cal P}_{(\alpha, \beta)}(I)=
\frac{\displaystyle 2}{\displaystyle \alpha \beta}
\int\limits^{\infty}_{0} dK e^{K/2}. J_0(2\sqrt{IK})\/
\int\limits^{\infty}_{0} dL\; 
{\displaystyle e}^{-L(\frac{\displaystyle 1}
{\displaystyle \alpha^2} + \frac{\displaystyle 1}{\displaystyle
\beta^2})}
J_0(2\sqrt{LK})\/I_0\/\left( L(\frac{\displaystyle 1}
{\displaystyle \alpha^2} - \frac{\displaystyle 1}{\displaystyle
\beta^2})\right) 
\label{PI-gauss-1}
\eeq
The first integral, over L, always converges thanks to the
asymptotic behaviours of $J_{0}(z)\/$ and $I_{0}(z)\/$:
\beqa
J_0(z) &\begin{array}{c}\longrightarrow \\ z 
\rightarrow + \infty \end{array}& \sqrt{\frac{\displaystyle
2}{\displaystyle \pi z}}\/ \cos{(z-\pi/4)},\nonumber \\
I_0(z) &\begin{array}{c}\longrightarrow \\ z 
\rightarrow + \infty \end{array}& \frac{\displaystyle e^{z}}
{\sqrt{\displaystyle 2\/\pi z}}
\label{asymptotic-I0}
\eeqa
Moreover, by suitable and permitted analytic continuation
of a standard definite integral available in the 
literature~[reference 16 p.711, formula 6.644]
we obtain a formula with whose help the L-integral can
be done explicitly. The requisite formula is, for real
parameters $a,b,c\/$ obeying $a > |c| \geq 0\/,b>0\/$:
\beq
\int \limits^{\infty}_{0} dx \; e^{-ax} \; J_0(2 \sqrt{bx})
\; I_0(cx) =\frac{\displaystyle 1}{\displaystyle \sqrt{a^2
-c^2}} \; \mbox{exp}\left(\frac{\displaystyle -ab}
{\displaystyle a^2 -c^2} \right) \; I_0
\left(\frac{\displaystyle cb}{\displaystyle a^2-c^2}\right)
\label{integral}
\eeq
Taking $a=\frac{\displaystyle 1}
{\displaystyle \alpha^{2}}+\frac{\displaystyle 1}
{\displaystyle \beta^{2}},
b=K,c= \frac{\displaystyle 1}
{\displaystyle \alpha^{2}}-\frac{\displaystyle 1}
{\displaystyle \beta^{2}}\/$ here 
and using the result in
eq~(\ref{PI-gauss-1}) we get for ${\cal P}_{(\alpha,\beta)}(I)\/$ the
single integral
\beqa
{\cal P}_{(\alpha, \beta)}(I)=
\int\limits^{\infty}_{0} dK\; e^{K/2}.\; J_0(2\sqrt{IK})\;
{\displaystyle e}^{-K (\frac{\displaystyle \alpha^2 + 
\beta^2}{\displaystyle 4})} 
\;I_0\left( \frac{\displaystyle K}{\displaystyle 4}(\alpha^2 -
\beta^2) \right) 
\label{PI-gauss-int}
\eeqa

First let us look at the classical nonsqueezed situation. Leaving
aside the marginal cases when $\alpha$ or $\beta$ equals
unity, we again use the result~(\ref{integral}) 
to evaluate~(\ref{PI-gauss-int}) explicitly:
\beqa
\alpha, \beta >1\quad : \quad 
{\cal P}_{(\alpha, \beta)}(I) =
2\/(\alpha^2-1)^{-1/2}.(\beta^2-1)^{-1/2}
\mbox{exp} \left[-I\left( \frac{\displaystyle 1}{\displaystyle
\alpha^2-1} + \frac{\displaystyle 1}{\displaystyle
\beta^2-1}\right)   \right]\;
I_0\;\left[I\left( \frac{\displaystyle 1}{\displaystyle
\alpha^2-1} - \frac{\displaystyle 1}{\displaystyle
\beta^2-1}\right)   \right]
\eeqa
This is explicitly nonnegative, and is consistent with the state
being classical. In this case, we can go further and obtain a
closed form expression for the photon-number probabilities 
$p_{(\alpha, \beta)}(n)\/$. We have:
\beqa
p_{(\alpha, \beta)}(n)&=&
\int\limits^{\infty}_{0} dI\/ {\cal P}_{(\alpha, \beta)}(I) 
{\displaystyle e}^{-I}\;\frac{\displaystyle I^n}{\displaystyle n!}
\nonumber \\
&=&\frac{\displaystyle 1}{\displaystyle n!}\;
\frac{\displaystyle 2}{\displaystyle \sqrt{(\alpha^2 -1)
(\beta^2-1)}} 
\int \limits^{\infty}_{0} dI\/{\displaystyle e}^{-a I}
\/I^n\;I_0(bI),
\nonumber \\
a&=&1+\frac{\displaystyle 1}{\displaystyle \alpha^2 -1}+
\frac{\displaystyle 1}{\displaystyle \beta^2 -1}
=\frac{\displaystyle \alpha^2\beta^2 -1}
{\displaystyle (\alpha^2 -1)(\beta^2 -1)},
\nonumber \\
b&=&\frac{\displaystyle (\beta^2-\alpha^2) }
{\displaystyle (\alpha^2 -1)(\beta^2 -1)}
\eeqa
The resulting integral is a known one leading to an expression
in terms of the hypergeometric function~[reference 16, p.711, formula
6.621]
\beq
\int \limits^{\infty}_{0} dx\; {\displaystyle e}^{-ax}\;
x^n\; I_0(bx)=
\frac{\displaystyle n!}{\displaystyle a^{n+1} }
F \left(\frac{\displaystyle n}{\displaystyle 2}+
\frac{\displaystyle 1}{\displaystyle 2},
\frac{\displaystyle n}{\displaystyle 2}+1;
1;
\frac{\displaystyle b^2}{\displaystyle a^2 } \right)
\eeq
so the probabilities $p_{(\alpha, \beta)}(n)\/$ are:
\beqa
p_{(\alpha, \beta)}(n)&=&
\frac{\displaystyle 2}{\displaystyle 
\sqrt{(\alpha^2-1)(\beta^2-1)}}. 
\left[\frac{\displaystyle (\alpha^2-1)(\beta^2-1)}
{\displaystyle \alpha^2\/\beta^2\/-1} \right]^{n+1}
F \left(\frac{\displaystyle n}{\displaystyle 2}+
\frac{\displaystyle 1}{\displaystyle 2},
\frac{\displaystyle n}{\displaystyle 2}+1;
1;z \frac{}{}
\right),
\nonumber \\
z&=&\left( \frac{\displaystyle \alpha^2 - \beta^2}
{\displaystyle \alpha^2 \/ \beta^2 \/-1} \right)^2 \quad,\quad
\alpha, \beta >1.
\label{PND-closed}
\eeqa
The combination $z\/$ of $\alpha\/$ and $\beta\/$ does not exceed unity
as we have $\alpha , \; \beta > 1\/$:
\beq
1-z = (\alpha^4-1)(\beta^4-1)/(\alpha^2 \beta^2 -1)^2.
\label{factor}
\eeq  
It is interesting to note that the result~(\ref{PND-closed}) for 
$p_{(\alpha ,
\beta)}(n)\/$ is a manifestly nonnegative closed-form expression;
in this respect it may be contrasted with the expression given
earlier in the literature~\cite{process-gaussian}
 \par
Next let us consider the squeezed regime $\beta < 1 ,\alpha
\geq 1/\beta\/$. Then the exponential factor $e^{K/2}$ in the
integral in eq~(\ref{PI-gauss-int}) overpowers the remaining factors:
\beqa
{\displaystyle e}^{K/2}\;{\displaystyle
e}^{-K\/(\alpha^2+\beta^2)/4}\; 
I_0\/ \left(K\/(\alpha^2-\beta^2)/4 \right)
\begin{array}{c}\longrightarrow \\ K 
\rightarrow + \infty \end{array}
\quad \frac{\displaystyle 1}{\displaystyle \sqrt{\alpha^2 -\beta^2}} 
\sqrt{\frac{\displaystyle 2}{\displaystyle \pi K}}
{\displaystyle e}^{K\/(1 - \beta^2)/2}
\eeqa
This means that ${\cal P}_{(\alpha,\beta)}(I)\/$ is no longer
the Fourier-Bessel transform of a square integrable function
of $K\/$; it has switched abruptly from being a classical
probability density for intensity to being a distribution, 
essentially as singular as is permitted by the general
considerations of the previous Section!
\par
There is thus no regime in which ${\cal P}_{(\alpha,\beta)}(I)\/$
remains "classical" while $\phi$ is not - the weakly nonclassical
possibility is not realised at all in the family of
Gaussian-Wigner states.
Even though ${\cal P}_{(\alpha, \beta)}(I)\/$ is a distribution in the 
squeezed regime, we can obtain the photon number probabilities by
analytic continuation starting from the result~(\ref{PND-closed}) in the
nonsqueezed case. The justification is the following. At the level of
Wigner distributions we know that the probability $p_{(\alpha,
\beta)}(n)\/$ is the phase space integral of the product of
$W_{(\alpha, \beta)}(q,p)\/$ and the Wigner function $W^{(n)}(q,p)\/$ for
the $n$th state of the harmonic oscillator~\cite{phys-rep-wigner}:
\beqa
\hat{\rho}&=&\vert n \rangle \langle n \vert \Rightarrow 
W^{(n)}(q,p)=\frac{\displaystyle (-1)^n}{\displaystyle \pi}
{\displaystyle e}^{\displaystyle -(q^2+p^2)}\/L_n
\left(2 (q^2+p^2) \right);
\nonumber \\
p_{(\alpha, \beta)}(n)&=& 2\/\pi \int \int dq\/dp \/ W_{(\alpha,
\beta)}(q,p) \/ W^{(n)}(q,p).
\eeqa
Here $L_n(.)\/$ is the $n\/$th order Laguerre polynomial. Using the
rotational invariance of $W^{(n)}(q,p)\/$ and
eq.~(\ref{wigner-angle-av}) for the angular
average of $W_{(\alpha, \beta)}(q,p)$, we can reduce $p_{(\alpha
,\beta)}(n)\/$  to a single radial phase space integral:
\beq
p_{(\alpha, \beta)}(n) = \frac{\displaystyle (-1)^n}{\displaystyle
\pi} \frac{\displaystyle 2}{\displaystyle \alpha \beta}\/ 2 \pi\/
\int \limits_{0}^{\infty}dL\/ \exp\left\{-2 L -L\left(\frac{1}{\alpha^2}+
\frac{1}{\beta^2} \right) \right\} L_n(4 L)\/ I_0\left(L(\frac{1}{\beta^2}-
\frac{1}{\alpha^2})\right)
\label{PND-wigner}
\eeq
This is valid for all $\alpha\/$ and $\beta\/$ subject to the
standard restrictions $\alpha, \; \beta >1\, \quad \alpha\/ \beta
\geq 1\/$. Since we have symmetry in $\alpha\/$ and $\beta\/$, we may
assume with no loss of generality that $\alpha \geq \beta\/$. Then
the asymptotic behaviour~(\ref{asymptotic-I0}) 
for $I_0(z)\/$ as $z \rightarrow
\infty\/$ shows that for large $L\/$ the integrand here behaves like
\beq
L^{n-1/2} \exp \left\{-2L\/(1+1/\alpha^2) \right\}
\eeq
Thus the integral~(\ref{PND-wigner}) is absolutely convergent for all
$\alpha\/$ and $\beta\/$, and is in fact analytic in these variables
(in the appropriate regions of the complex planes).

Having established this, we may now go back to the closed expression 
~(\ref{PND-closed}) valid in the nonsqueezed case and analytically continue it
to $\beta < 1\/$, $\alpha \/ \beta\geq 1\/$. Now from eq.~(\ref{factor}) we
see that the argument $z\/$ of the hypergeometric function exceeds
unity, which lies outside the domain of convergence of the power
series expansion of $F(\frac{n+1}{2},\frac{n}{2}+1;1;z)$. By analytically
continuing to $z > 1\/$, and keeping track of phases generated in
switching from $(\beta^2-1)\/$ to $(1- \beta^2)\/$ in the prefactors
in eq.~(\ref{PND-closed}), we find that in the squeezed regime we have
different expressions for $p_{(\alpha, \beta)}(n)\/$ for even $n\/$
and for odd $n\/$: 
\beqa
p_{(\alpha, \beta)}(n)&=&
\frac{\displaystyle 2 }{\displaystyle \sqrt{\pi}}
\frac{\displaystyle [(\alpha^2-1)\/(1-\beta^2)]^{n+1/2}}
{\displaystyle (\alpha^2\/\beta^2-1)^{n+1}}
\frac{\displaystyle 1}{\displaystyle z^{\frac{n+1}{2}}}
\left\{ \begin{array}{c} 
\frac{\displaystyle \Gamma(m+1/2)}{\displaystyle m!}
F\left(m+1/2,m+1/2;1/2;\frac{\displaystyle 1}{\displaystyle z} \right)
,\;n=2m,
\\ 
\frac{\displaystyle 2}{\displaystyle \sqrt{z}}
\frac{\displaystyle \Gamma(m+3/2)}{\displaystyle m!}
F\left(m+3/2,m+3/2;3/2;\frac{\displaystyle 1}{\displaystyle z} \right)
,\;n=2m+1,
\end{array} \right.
\nonumber \\
z&=& \left(\frac{\displaystyle \alpha^2 -\beta^2}
{\displaystyle \alpha^2\/\beta^2 -1} \right)^2 \quad >1, \quad \alpha > 
\frac{\displaystyle 1}{\displaystyle \beta},\quad \beta < 1.
\label{PND-closed-I}
\eeqa
Once again we have manifestly nonnegative closed form
expressions~\cite{process-gaussian}

The actual expressions for the first few probabilities show the
general trend. We find after simplification that, as expected both
eq.~(\ref{PND-closed}) and eq.~(\ref{PND-closed-I}) 
give identical functions of $\alpha\/$
and $\beta\/$:
\beqa
p_{(\alpha, \beta)}(0) &=& 
2 \times \left\{(\alpha^2+1)(\beta^2+1)\right\}^{-1/2}
\nonumber \\
p_{(\alpha, \beta)}(1) &=& 
2\/(\alpha^2\/ \beta^2-1) \times \left\{(\alpha^2+1)(\beta^2+1)\right\}^{-3/2}
\nonumber \\
p_{(\alpha, \beta)}(2) &=& 
 \left\{(\alpha^2-\beta^2)^2\/+\/ 2\/(\alpha^2\/ \beta^2-1)^2\right\}
\times \left\{(\alpha^2+1)(\beta^2+1)\right\}^{-5/2}
\nonumber \\
p_{(\alpha, \beta)}(3) &=& (\alpha^2\/
\beta^2-1)\/\left\{3\/(\alpha^2-\beta^2)^2\/+\/2\/
(\alpha^2\/ \beta^2-1)^2\right\}
\times \left\{(\alpha^2+1)(\beta^2+1)\right\}^{-7/2}
\nonumber \\
p_{(\alpha, \beta)}(4) &=& \frac{1}{4}
\left\{3\/(\alpha^2-\beta^2)^4\/+\/24\/
(\alpha^2-\beta^2)^2\/(\alpha^2\/ \beta^2-1)^2+
8\/(\alpha^2\/ \beta^2-1)^4 \right\}
\times \left\{(\alpha^2+1)(\beta^2+1)\right\}^{-9/2}
\nonumber \\
p_{(\alpha, \beta)}(5) &=& \frac{1}{4}
(\alpha^2\/
\beta^2-1)\/\left\{15\/(\alpha^2-\beta^2)^4\/+\/40\/
(\alpha^2-\beta^2)^2\/(\alpha^2\/ \beta^2-1)^2+
8\/(\alpha^2\/ \beta^2-1)^4 \right\}
\times \left\{(\alpha^2+1)(\beta^2+1)\right\}^{-11/2}
\label{pn-1234}
\eeqa
The appearance of the ``uncertainty principle factor'' 
$(\alpha^2 \beta^2 -1)\/$
in $p_{(\alpha,\beta)}(n) \/$ for odd $n\/$ alone is immediately
understandable: when the uncertainty limit is saturated and $\alpha
\beta =1\/$ , the Gaussian Wigner function $W_{(\alpha,
1/\alpha)}(q,p)\/$ describes the squeezed vacuum, for which it is well
known that  $p_{(\alpha,1/ \alpha)}(n)\/$ vanishes when $n\/$ is
odd~\cite{schliech}. Conversely, even in the nonsqueezed regime, despite the
uniform looking expression~(\ref{PND-closed}), there is a discrimination
between the cases of even and odd $n\/$ which is seen
when the hypergeometric function is worked out in detail. In the
limit $\alpha = \beta =1$, we have of course just the vacuum state,
and then $p_{(1,1)}(n)\/$ vanishes for all $n \geq 1\/$. This case be
seen quite explicitly in the expressions displayed in eq.~(\ref{pn-1234}).
\section{Concluding Remarks}
We have examined the class of Gaussian-Wigner distributions
for a single mode radiation field in quantum optics from the
point of view of a recently introduced classification of
quantum states into three mutually exclusive types - 
classical, weakly nonclassical and strongly nonclassical. We
have found that only the first and third possibilities arise
in this case, corresponding respectively to the
nonsqueezed and squeezed situations. As shown elsewhere, there
is an interesting class of pure states which give physical
examples of the weakly nonclassical type. These are superpositions
of the number states of the following general type:
\beq
\vert \psi \rangle = {\displaystyle e}^{-\frac{\displaystyle 1}
{\displaystyle 2}\vert \displaystyle \alpha \vert^2}
\sum^{\infty}_{n=0}
\frac{\displaystyle \alpha^{n}}
{\displaystyle \sqrt{n!}} {\displaystyle e}^{\displaystyle i
\beta (n)} \vert n \rangle, 
\label{weakly-nonc-states}
\eeq
where $\alpha\/$ is any complex number and $\beta(n)\/$ is a
\underline{nonlinear} function of $n\/$. Here the 
photon number probabilities
are independent of $\beta(n)\/$  and follow the Poisson
distribution, so ${\cal P}_{\psi}(I)\/$
is a delta function:
\beqa
 {\cal P}_{\psi}(I) = \delta (I-\alpha^{\star}\alpha)
\eeqa
However on the basis of Hudson's Theorem~\cite{hudson-th} it turns out that
the Wigner function $W_{\psi}(q,p)\/$, which is 
\underline{not} Gaussian, must be negative somewhere, so in
turn $\phi(z)$ cannot be nonnegative. This shows that
the states~(\ref{weakly-nonc-states}) are weakly nonclassical.
\par
Our result that the centered Gaussian-Wigner distributions are never
weakly nonclassical has an important physical consequence. In the
regime $\alpha > 1, \beta < 1\/$ which corresponds to {\em quadrature}
squeezing, since ${\cal P}(I)\/$ is not nonnegative the nonclassical
nature of the state {\em must already show up} in properties of the
photon number distribution probabilities $p_{(\alpha, \beta)}(n)\/$,
ie., via phase insensitive quantities. The simplest such signal,
namely subpoissonian statistics, does not however display the
nonclassicality of the state~\cite{process-gaussian}. We find after
simple algebra that the Mandel Q-parameter is   always nonnegative:
\beqa
Q(\alpha, \beta)&=& 
\frac{\displaystyle \langle \hat{a}^{\dagger^2} \hat{a}^2\rangle - 
\langle \hat{a}^{\dagger} 
\hat{a}\rangle^2 }{\displaystyle \langle \hat{a}^\dagger \hat{a}\rangle }
\nonumber \\
&&=2\/ \left\{ (\alpha^2-1)^2+(\beta^2-1)^2\right\}/
 (\alpha^2 + \beta^2-2)^2 \geq 0 
\eeqa
There are however (infinitely many) other signatures of a
nonclassical photon number distribution, some of which are local in
that they involve only a few contiguous probabilities $p(n)\/$. For
example we have the result~\cite{characterization}:
\beqa
{\cal P}(I) \geq 0 \Rightarrow \quad l(n)\;&=
&(n+1)\/p(n-1)\/p(n+1)-n\/(p((n))^2 \geq 0,\nonumber \\
&& n=1,2,3,...
\eeqa
Therefore if any $l(n)\/$ is negative for some given state, that is
evidence for the strongly nonclassical nature of that state.
For the states $W_{(\alpha, \beta)}(q,p)\/$, taking $\alpha=2\/,
\;\frac{1}{2} < \beta < 1\/$ as an example, we do find explicitly as
shown in Figure~1 that
$l(2),\;l(4),\;l(6)\cdots\/$ are negative for some range of values of
$\beta\/$ before turning positive as $\beta \/$ increases; while
$l(1),\;l(3),\;l(5)\cdots\/$ do not display such nonclassical
behaviour. 
\begin{figure}
\hspace{3.5cm} \epsfxsize=10cm \epsfbox{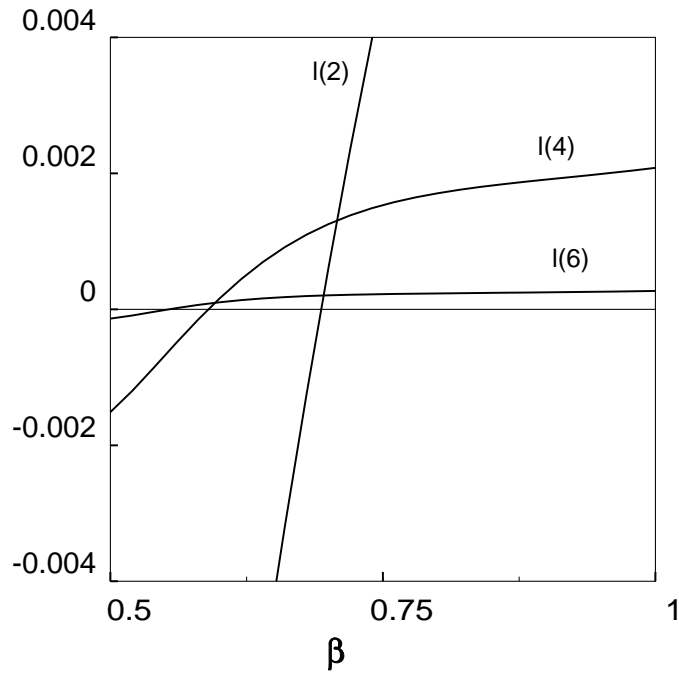}
\caption{Violation of the local conditions on photon number
distribution in the squeezed regime}
\end{figure}

It is expected that our conclusions will not be altered
drastically if we consider general noncentred Gaussian-Wigner
distributions. This aspect and other examples of states and the
cases of two or more modes, will be taken up elsewhere.


\begin{references}
\bibitem[\star]{email} email arvind@physics.iisc.ernet.in
\bibitem[\dag]{jncasr} Also at Jawaharlal Nehru Centre for 
               Advanced Scientific 
               Research, Jakkur, Bangalore - 560 064, India.
\bibitem {walls-nature} M.~C.~Teich and B.~E.~A.~Saleh in
{\it Progress in Optics\/}, Vol. 26, 
ed. E.~Wolf (North-Holland, Amsterdam, 1988);
D.~F.~Walls, Nature {\bf 280}, 451 (1979).
\bibitem {squeezing-1}
D.~Stoler, Phys. Rev. D {\bf 1}, 3217 (1970);
H.~P.~Yuen, Phys. Rev. A {\bf 13}, 2226 (1976);
D.~F.~Walls, Nature {\bf 306}, 141 (1983)
\bibitem {antibunching}
H.~J.~Kimble, M.~Dagenais and L.~Mandel, 
Phys. Rev. Letters {\bf 39} 691 (1977). 
\bibitem {sub-poissonian}
 R.~Short and L.~Mandel, Phys. Rev. Letters 
{\bf 51}, 384 (1983).
\bibitem {others} C.~K.~Hong and L.~Mandel, Phys. Rev.
A {\bf 32}, 974 (1985).
G.~S.~Agarwal and K.~Tara, Phys.
Rev. A {\bf 46} 485 (1992).
\bibitem{diag-coh} 
R.~J.~Glauber, Phys. Rev. {\bf 131}, 2766 (1963);
E.~C.~G.~Sudarshan, Phys Rev. Lett. {\bf 10}, 277 (1963).
\bibitem {characterization}
Arvind, N.~Mukunda and R.~Simon, 
{\it Characterisations of Classical and Non-classical 
         states of Quantised Radiation}
IISc Preprint (1996).
\bibitem {process-gaussian}
G.~S.~Agarwal, J. Mod. Opt. {\bf 34}, 909 (1987);
 G.~S.~Agarwal and G.~Adam Phys. Rev. A {\bf 38}, 750 (1988);
S.~Chaturvedi and V.~Srinivasan Phys. Rev. A {\bf 40}, 6095 (1989).
\bibitem {distributions}
J.~R.~Klauder and E.~C.~G.~Sudarshan, 
{\it Fundamentals of Quantum Optics\/}, Benjamin, New York
(1968); I.~M.~Gel'fand and G.~E.~Shilov, {\it Generalised
functions, Vol.I Properties and Operations}, Academic Press, New
York (1964).
\bibitem {weyl-book}
H.~Weyl, {\it The Theory of Groups and Quantum Mechanics}
(Dover, New York, 1931), p.275.
\bibitem {wigner-1932} 
E.~P.~Wigner, Phys. Rev. A {\bf 40}, 749 (1932).
\bibitem {forier-bessel}
N.~N.~Lebedev, {\it Special functions and their applications},
(Dover New York, 1972), p.130. 
\bibitem {gaussian-wigner-1}
R.~G.~Littlejohn, Phys. Rep. {\bf 138}, 193 (1986).
\bibitem {gaussian-wigner-2}
R.~Simon, N.~Mukunda and E.~C.~G.~Sudarshan,
Phys. Lett. A {\bf 124} 223 (1987).
R.~Simon, N.~Mukunda and E.~C.~G.~Sudarshan,
Phys. Rev. A {\bf 36} 3868 (1987); R.~Simon, N.~Mukunda and B.~Dutta, 
Physical Review A {\bf 49}, 1567 (1994).
\bibitem {squeezing}
B.~Dutta, N.~Mukunda, R.~Simon and A.~Subramaniam,
Journal of the Optical Society of America B {\bf 10}, 253 (1993);
Arvind, Biswadeb Dutta, N. Mukunda and 
R.~Simon,  Pramana Jr. of Physics  {\bf 45}, 471 1995.
\bibitem {integral}
I.~S.~Gradshteyn and I.~W.~Ryzhik, {\it Tables of Integrals,
Series and Products} (Academic Press NY 1972). 
\bibitem {phys-rep-wigner} M.~Hillery et. al., Phys. Reps.
{\bf 106}, 121 (1984).
\bibitem {schliech} W.~Schleich and J.~A.~Wheeler, Nature {326}, 574 (1987);
W.~Schleich and J.~A.~Wheeler, J. Opt. Soc Am B
{\bf 4}, 1715 (1990). 
\bibitem {hudson-th} R.~L.~Hudson, Rep. Math. Phys. {\bf
6}, 249 (1974). 
\end{references}
\end{document}